%% file: main.tex
\documentclass[11pt]{article}

\usepackage[a4paper, total={6in, 8in}]{geometry}
\usepackage{afterpage}
\usepackage{amsmath}
\usepackage{amssymb}
\usepackage{amsthm}
\usepackage{bashful}
\usepackage[utf8]{inputenc}
\usepackage{csquotes}
\usepackage{graphicx}
\usepackage{mathtools}
\usepackage{mathrsfs}
\usepackage[english]{babel}
\usepackage{graphicx}
\usepackage{enumerate}
\usepackage{soul}
\usepackage{xcolor}
\usepackage{url} 

\usepackage[ backend=biber, style=numeric, citestyle=numeric, sorting=none]{biblatex}
\DeclareSymbolFont{lettersA}{U}{txmia}{m}{it}
\DeclareMathSymbol{\phiup}{\mathbb}{lettersA}{20}

\DeclareMathAlphabet{\mymathbb}{U}{bbold}{m}{n}

	\newcommand{\blind}{0}
	
    \makeatletter
    \renewcommand\section{\@startsection {section}{1}{\z@}%
                                       {-3.5ex \@plus -1ex \@minus -.2ex}%
                                       {2.3ex \@plus.2ex}%
                                       {\normalfont\fontsize{16}{19}\bfseries}}
    \renewcommand\subsection{\@startsection{subsection}{2}{\z@}%
                                         {-3.25ex\@plus -1ex \@minus -.2ex}%
                                         {1.5ex \@plus .2ex}%
                                         {\normalfont\fontsize{14}{17}\bfseries}}
    \renewcommand\subsubsection{\@startsection{subsubsection}{3}{\z@}%
                                        {-3.25ex\@plus -1ex \@minus -.2ex}%
                                         {1.5ex \@plus .2ex}%
                                         {\normalfont\normalsize\fontsize{14}{17}\selectfont}}
    \makeatother
	
    \usepackage{blindtext} 
    \usepackage{setspace}
    \usepackage[symbol]{footmisc}

\begin{document}
		\def\spacingset#1{\renewcommand{\baselinestretch}%
			{#1}\small\normalsize} \spacingset{1}
		
		\if0\blind
		{
			\title{\bf Thermodynamics of 5D Charged Rotating Black Holes:\\ A Counterterms Treatment}
			\author{Adel Awad$^{\, a,b}$ and Hassan ElSayed$^{\, c,d,}$\footnote{Corresponding author; hassan.elsayed@uconn.edu} \\ 
            \quad \\
			$^a$ Department of Physics, Faculty of Science, Ain Shams University, Cairo 11566, Egypt \\
             $^b$ Centre for Theoretical Physics, The British University in Egypt, P.O. Box 43, \\ El Sherouk City, Cairo 11837, Egypt\\
             $^c$  Department of Physics, University Connecticut, Storrs, CT 06269-3046, USA\\
             $^d$ Department of Physics, School of Sciences and Engineering, American University in Cairo, \\P.O. Box 74, AUC Avenue New Cairo, Cairo, Egypt}
			\date{}
			\maketitle
		} \fi
		
		\if1\blind
		{

            \title{\bf \emph{IISE Transactions} \LaTeX \ Template}
			\author{Author information is purposely removed for double-blind review}
			
\bigskip
			\bigskip
			\bigskip
			\begin{center}
				{\LARGE\bf \emph{IISE Transactions} \LaTeX \ Template}
			\end{center}
			\medskip
		} \fi
		\bigskip

\begin{abstract}
    We use the counterterms subtraction method to calculate various thermodynamical quantities for charged rotating black holes in five-dimensional minimal gauged supergravity \cite{pope}. Specifically, we analyze certain issues related to the first law and Smarr's relation in the presence of a conformal anomaly. Among the bulk quantities calculated are the on-shell action, total mass, and angular momenta of the solution. All these quantities are consistent with previous calculations made using other methods. For the boundary theory, we calculate the renormalized stress tensor, conformal anomaly, and Casimir energy. Using the Papadimitriou-Skenderis analysis \cite{skenderis}, we show that the mass calculated via the counterterms method satisfies the first law of black hole thermodynamics. To discuss extended thermodynamics, we extend the definition of the thermodynamic volume to cases with conformal anomalies using a procedure similar to that of Papadimitriou-Skenderis. We show that this volume correctly accounts for extra terms due to boundary metric variation. This shows that the mass and volume calculated using counterterms satisfy Smarr's relation as well as the first law.
\end{abstract}
	 		
	\noindent%

	\spacingset{1.5} 

\tableofcontents
\input{1-Introduction}
\input{2-Presenting_the_Solution}

\input{3-Thermodynamics_with_M0}
\input{4-Counterterms_and_the_First_Law}

\input{5-Holography}
\input{6-Thermodynamics_with_M}
\input{7-Conclusion}


\if0\blind{
\section*{Acknowledgments}
This work was partially supported by research support grant RSG1-2018-FY18-FY19 from the American University in Cairo; id:10.13039/501100009229.	} \fi

\spacingset{1}
\printbibliography[title={References}]
	
\end{document}

%% file: 1-Introduction.tex
\section{Introduction} \label{sec:introduction}

The discovery of the anti-de sitter/conformal field theory (AdS/CFT) correspondence \cite{large_n_lim} invoked lots of interest in asymptotically AdS solutions in general relativity. This duality reveals how various gravitational solutions in the AdS side encode important information about a specific gauge field theory on the boundary in their semi-classical gravitational action. A strong interest in studying black hole solutions in AdS$_5$ spacetimes then followed after studying a specific example of this duality, namely, the equivalence between string theory on AdS$_5\times S^5$ and four-dimensional ${\cal N}=4$, Super Yang-Mills (SYM) theory on the boundary. This duality and other similar ones allow one to map certain five-dimensional AdS solutions to certain states in the boundary field theory. Furthermore, it shows that bulk boundary quantities work as sources to CFT quantities, e.g., the boundary metric works as a source for the boundary energy-momentum tensor. In this duality as well as in a Euclidean path-integral formalism we have certain boundary conditions which fix bulk boundary quantities. For example, AdS black hole thermodynamics with its possible phase transitions are investigated for a fixed boundary metric as is discussed in \cite{Hawking:1982dh,witten1998}.

\par
The AdS/CFT correspondence relates a strong-coupling regime in the boundary with a weak-coupling regime in the bulk. The bulk partition function can be approximated \cite{boruch} using saddle points leading to

\begin{equation}
    \mathcal{Z}_\text{grav} \simeq \exp(-I),
\end{equation}

where $I$ is the on-shell gravitational action. However, in AdS spacetimes, the action as a volume integral diverges as we  take the radial coordinate $r$ to infinity. There are two main techniques often used to regulate this action: the background subtraction method (for example, see\cite{gibbons2005} and references therein) and the counterterms subtraction method \cite{surface_terms}. Background subtraction works by defining a reference background spacetime, then calculating the finite action as the difference between the action of the spacetime and that of the background metric. Since both metrics have the same diverging asymptotic region, this subtraction method leaves a finite action. But a disadvantage of this technique is that the choice of a proper background reference is not always obvious or clear \cite{holography_and_weyl_anomaly,kraus,surface_terms}. In addition, sometimes certain quantities which are common between the spacetime and its background (such as the conformal anomaly and vacuum energy) cancel out in this procedure \cite{awad_johnsson1}.

\par 

The counterterms subtraction method \cite{holography_and_weyl_anomaly,kraus,surface_terms} works by adding certain covariant boundary counterterms to the action which cancel action divergences exactly. These counterterms are thus inherent to the spacetime boundary and there is no need for a background or a reference spacetime. Moreover, the employment of the counterterms subtraction technique allows one to compare some important quantities between the bulk and the boundary. For instance, the trace of the Brown-York quasilocal stress tensor \cite{brown_york} resulting from the action of the counterterms technique is related to the conformal anomaly on the boundary. Another example is the total energy/mass found using the counterterms technique, which includes a non-vanishing contribution as the mass parameter $m$ is sent to zero. In the context of the AdS/CFT correspondence, this contribution which is the  spacetime background energy is interpreted as the Casimir energy of the boundary CFT. These two phenomena appear particularly in odd-dimensional bulk theories, or even-dimensional boundary field theories. For example, see \cite{kraus,awad_johnsson1} and references therein.

\par 

The counterterms method was criticized in \cite{gibbons2005} where the authors argued that the first law of black hole thermodynamics,

\begin{equation} \label{first_law1}
    dM = TdS + \sum_i \Omega_i dJ_i + \Phi dQ,
\end{equation}

is not satisfied through this procedure. Several authors have presented resolutions to this apparent violation \cite{awad2007, boruch, skenderis}. Here we are particularly interested in the resolution presented in \cite{skenderis}, which we briefly explain below.

\par 
It is well known that quantum corrections can break classical conformal symmetries in even-dimensional field theories leading to conformal anomalies. The conformal or Weyl transformation on the boundary is a subset of the bulk diffeomorphisms. This diffeomorphism is a Penrose-Brown-Hennaux (PBH) transformation \cite{Penrose:1986ca,Brown:1986nw}. In the case of a non-vanishing conformal anomaly, a PBH transformation will not leave the gravitational on-shell action invariant. The action has a relation to other quantities through the Gibbs-Duhem given by
\begin{equation} \label{GD}
    I=\beta {M} - TS - \sum_i \Omega_i J_i - \Phi Q.
\end{equation}
Now by varying this equation, one is expected to get variations of all terms, but instead, we get
\begin{equation} \label{vGD}
    \delta I=\beta \delta {M}.
\end{equation}
In other words, this variation affects only the mass, but not the other quantities in the first law, since entropy, electric charge $Q$, and angular momenta $J_a$ and $J_b$ can be expressed in terms of integrals over the horizon \cite{skenderis}. 
In $\S$\ref{sec:calculation_of_M_and_J} we show that the counterterms mass is the sum of two terms, the first is localized in the black hole region (depends on the mass parameter and electric charge) while the second term is the background energy of the spacetime which is not restricted to the black hole region. The flux of the latter term depends on the boundary metric. An important boundary condition for the variation problem \cite{skenderis} is to keep the boundary metric fixed by utilizing a PBH or boundary Weyl transformation. Therefore, one can use a Weyl transformation to cancel the boundary metric variation of the mass term in the first law. As a result, the first law is satisfied.
To summarize, the first law is not violated by the counterterms mass, but one should be careful about how to compute variations of bulk quantities since those must be done at a fixed boundary metric as the AdS/CFT requires.


\par

In this work, we use the counterterms subtraction method to calculate the renormalized on-shell action for the general rotating charged AdS solution presented in \cite{pope} as well as its mass and angular momenta. We show that these quantities satisfy the first law and the Gibbs-Duhem relation. Going to extended thermodynamics \cite{Kastor:2009wy,Kubiznak:2012wp} where we allow the cosmological constant to vary and act as a pressure, a naive calculation of the first law shows that it is not satisfied. We show that a similar issue exists with the volume defined in extended thermodynamics \cite{Kastor:2009wy,Kubiznak:2012wp}. Meaning that varying this quantity does not leave the boundary metric fixed and one needs to use a compensating PBH term to keep the volume fixed. This modification is important to satisfy the first law as well as the generalized Smarr's formula in counterterms context with a conformal anomaly.



The rest of the paper is organized as follows: in $\S$\ref{sec:presentation_and_discussion} we present the black hole solution and its thermodynamic quantities as in \cite{pope}. In $\S$\ref{sec:thermodynamics} we show that these quantities satisfy the standard thermodynamic relations: the first law \eqref{first_law1} and the Gibbs-Duhem relation. In $\S$\ref{sec:CTs_and_the_first_law} we use the counterterms subtraction method to calculate finite expressions for the action, mass, and angular momenta of the solution. In $\S$\ref{sec:holography} we calculate the renormalized stress tensor and conformal anomaly of the CFT from the dual gravitational theory and compare those results to the field theory calculations on rotating Einstein Universe. Also, we calculate the Casimir energy of the CFT and compare it to the background energy of the bulk theory which was calculated using the counterterms subtraction method. In $\S$\ref{sec:thermodynamics_with_CTs} we show that the quantities calculated using counterterms subtraction satisfy the first law in regular and extended thermodynamics. We study the effect of a Weyl transformation on the boundary and use this to show that the mass calculated using the counterterms method satisfies the first law of thermodynamics. Furthermore, we check that our expressions for action and mass satisfy the Gibbs-Duhem relation. Finally, we discuss the effect of boundary variations on Smarr's formula and propose a new modification to the thermodynamic volume in the presence of a conformal anomaly. This new volume satisfies the first law in extended phase-space as well as Smarr's formula.

%% file: 2-Presenting_the_Solution.tex
\section{The 5D Charged Rotating AdS Solution} \label{sec:presentation_and_discussion}

The Einstein-Hilbert-Chern-Simons Lagrangian in five-dimensions has the following form \cite{pope}

\begin{equation} \label{pope_lagrangian}
    \mathcal{L} = (R^2 + 12g^2) \star \mymathbb{1} - \frac{1}{2} \star F \wedge F + \frac{1}{3 \sqrt{3}} F \wedge F \wedge A.
\end{equation}

The first term on the right-hand side is the gravitational Einstein-Hilbert Lagrangian in AdS$_5$, the second is the Maxwell Lagrangian and the third is the Chern-Simons Lagrangian in five dimensions. The latter is required in five-dimensional gauged supergravity \cite{catastrophic_holography}.

In this letter we are interested in studying the general non-extremal rotating black holes in minimal five-dimensional gauged supergravity \cite{pope} which has the following metric:
\begin{equation} \label{metric}
    \begin{split}
       ds^2 = &-\frac{\Delta_\theta[(1+g^2r^2)\rho^2\text{d}t + 2q\nu]\text{d}t}{\Xi_a\Xi_b\rho^2} + \frac{2q\nu\omega}{\rho^2} + \frac{f}{\rho^4}\left( \frac{\Delta_\theta \text{d}t}{\Xi_a\Xi_b} - \omega \right)^2 + \frac{\rho^2\text{d}r^2}{\Delta_r} + \frac{\rho^2\text{d}\theta^2}  {\Delta_\theta}  \\ & + \frac{r^2+a^2}{\Xi_a}\sin^2\theta \,\text{d}\phi^2 
    	 +  \frac{r^2+b^2}{\Xi_b}\cos^2\theta \,\text{d}\psi^2,
    \end{split}
\end{equation}
where
\begin{align}
	& \nu=b\sin^2\theta \,\text{d}\phi+a \cos^2\theta \,\text{d}\psi, \quad \omega = a \sin^2\theta \frac{\text{d}\phi}{\Xi_a}+ b \cos^2\theta \frac{\text{d}\psi}{\Xi_b},\\
	&  \Delta_r = \frac{(r^2+a^2)(r^2+b^2)(1+g^2r^2)+q^2+2abq}{r^2}-2m, \label{Delta_r}\\
	&  \Delta_\theta = 1-a^2g^2\cos^2\theta - b^2g^2\sin^2\theta,\quad \rho^2 = r^2+a^2\cos^2\theta + b^2\sin^2\theta,\\
    &  \Xi_a = 1-a^2g^2, \quad \Xi_b = 1-b^2g^2,\\
	&  f=2m\rho^2-q^2+2abqg^2\rho^2.
\end{align}

The constant $g=1/\ell$, where $\ell$ is the AdS radius, is not to be confused with the determinant of the metric. 

\par 

The coordinate system in \eqref{metric} is $(t,r,\theta,\phi,\psi)$. The ranges of the last three coordinates is such that they cover a three-sphere. The range for $\theta$ is between $0$ and $\pi/2$, while that for $\phi$ and $\psi$ is between $0$ and $2\pi$. The electromagnetic four-potential is given in \cite{pope} by

\begin{equation}
    A = \frac{\sqrt{3}q}{\rho^2} \left( \frac{\Delta_\theta \text{d}t}{\Xi_a \Xi_b} - \omega \right).
\end{equation}

And the electric charge is given by

\begin{equation}
   Q = \frac{\sqrt{3} \pi  q}{4 \Xi _a \Xi _b}.
\end{equation}

In four spatial dimensions, the black hole has two possible rotation axes. The corresponding angular velocities are denoted $\Omega_a$ (in the $\phi$-direction) and $\Omega_b$ (in the $\psi$-direction). They are given in \cite{pope} by
\begin{align}
    \Omega_{a} &=\frac{a\left(r_{+}^{2}+b^{2}\right)\left(1+g^{2} r_{+}^{2}\right)+b q}{\left(r_{+}^{2}+a^{2}\right)\left(r_{+}^{2}+b^{2}\right)+a b q}, \quad \Omega_{b} =\frac{b\left(r_{+}^{2}+a^{2}\right)\left(1+g^{2} r_{+}^{2}\right)+a q}{\left(r_{+}^{2}+a^{2}\right)\left(r_{+}^{2}+b^{2}\right)+a b q}.
\end{align}
Evidently, the metric in \eqref{metric} has axial symmetries in $\phi$ and $\psi$. The Killing vectors associated with these symmetries are $\partial_\phi$ and $\partial_\psi$. The angular momenta were subsequently calculated in \cite{pope} using the Komar integral,

\begin{equation} \label{Ja/b_Komar}
    J_{a} = \frac{1}{16 \pi} \int_{S^3} \star \text{d} \left(\partial_{\phi}\right), \quad J_{b} = \frac{1}{16 \pi} \int_{S^3} \star \text{d} \left(\partial_{\psi}\right),
\end{equation}

yielding
\begin{align} \label{pope_angular_momenta}
    J_{a} &=\frac{\pi\left[2 a m+q b\left(1+a^{2} g^{2}\right)\right]}{4 \Xi_{a}^{2} \Xi_{b}}, \quad J_{b}=\frac{\pi\left[2 b m+q a\left(1+b^{2} g^{2}\right)\right]}{4 \Xi_{b}^{2} \Xi_{a}}.
\end{align}
Note that the integrals in \eqref{Ja/b_Komar} directly lead to finite results and do not need regularization.



%% file: 3-Thermodynamics_with_M0.tex
\section{Thermodynamics: Background Method} 
\label{sec:thermodynamics}

In this section we review the thermodynamics of the spacetime under consideration using background-subtraction calculations. For this solution, the temperature and entropy are given by
\begin{align}
    &T = \frac{r_+^4 \left[g^2 \left(a^2+b^2+2 r_+^2\right)+1\right]-(a b+q)^2}{2 \pi  r_+ \left[\left(a^2+r_+^2\right) \left(b^2+r_+^2\right)+a b q\right]}, \\
    &S = \frac{\pi^{2}\left[\left(r_{+}^{2}+a^{2}\right)\left(r_{+}^{2}+b^{2}\right)+a b q\right]}{2 \Xi_{a} \Xi_{b} r_{+}}.
\end{align}

 The total energy/mass was calculated in \cite{pope} by integration of the first law \eqref{first_law1}. This mass is given by\footnote{The mass was denoted in \cite{pope} by $M$ but we will use here the notation $M_0$ to distinguish this mass from that which we will derive using the counterterms method in $\S$\ref{sec:calculation_of_M_and_J}.}

\begin{equation} \label{M_0}
    M_0 = \frac{\pi  m \left(2 \Xi _a + 2 \Xi _b - \Xi _a\Xi _b\right) + 2 \pi  a b g^2 q \left(\Xi _a+\Xi _b\right) }{4 G \Xi _a^2 \Xi _b^2}.
\end{equation}

It is worth noting that this expression for the mass matches the one that was found using the ADM calculation in \cite{chen2005, boruch}. 
\par

Using the above mass, it is straightforward to check the validity of the first law 

\begin{equation} \label{first_law_vr1_pope_metric}
    dM_0 = T dS + \Omega_a dJ_a + \Omega_b dJ_b + \Phi dQ.
\end{equation}

The electric potential is found from
\begin{align} \label{Phi}
    \Phi &=  \xi^aA_a|_{r \to \infty} - \xi^aA_a|_{r \to r_+} \nonumber \\
    &=\frac{\sqrt{3} q r_+^2}{\left(a^2+r_+^2\right) \left(b^2+r_+^2\right)+a b q},
\end{align}

which is in agreement with \cite{grunau}.

\par 

To check the first law, one needs to verify the following equation

\begin{equation} \label{first_law_3_alpha}
    \frac{\partial M_0}{\partial\alpha}d\alpha= T \frac{\partial S}{\partial\alpha} d\alpha + \Phi\frac{\partial Q}{\partial\alpha} d\alpha + \Omega_a\frac{\partial J_a}{\partial\alpha} d\alpha + \Omega_b\frac{\partial J_b}{\partial\alpha} d\alpha,
\end{equation}

for the parameters, $\alpha = r_+, q, a, b$, since the mass is a function of these variables. Direct evaluation of these expressions shows that the first law is verified.

{\bf Gibbs Free Energy and Action Calculation}

 Since the integral of \eqref{pope_lagrangian} is divergent as $r \to \infty$, the background subtraction method instructs us to start by taking a cut-off at a large value $R$. We then specify a background metric by taking $m \to 0, q \to 0$ in \eqref{metric}, which gives the Kerr-AdS$_5$ spacetime. We compute the integral of this action at the cut-off value $R$, then subtract it from our previous calculation with the original metric to cancel the divergences. The action using the background subtraction method was given in \cite{chen2005} by 

\begin{equation} \label{I_bg_subtraction}
    I_0 = \frac{\pi  \beta }{4 \Xi_a \Xi_b}\left[m -\left(g^2 \left(a^2+r^2\right) \left(b^2+r^2\right)\right)-\frac{q^2 r^2}{\left(a^2+r^2\right) \left(b^2+r^2\right)+a b q}\right].
\end{equation}

It is straightforward to verify that
\begin{equation}
    I_0 = \beta G_0,
\end{equation}
where the Gibbs free energy is given by\footnote{We use the subscript ``0" to refer to the Gibbs free energy calculated from $M_0$ to distinguish it from that calculated from the counterterms mass.}

\begin{equation}
    G_0(T, \Omega_a,\Omega_b, \Phi) = M_0-T S-\Omega_a J_a-\Omega_b J_b-\Phi Q.
\end{equation}

This is the Gibbs-Duhem relation, one of the few important thermodynamic relations used to check the consistency of black hole thermodynamics. To check that $G_0=G_0(T, \Omega_a,\Omega_b, \Phi)$, one might vary $G_0$ to obtain
\begin{equation}
    dG_0=-SdT-J_a d\Omega_a-J_b d\Omega_b-Qd\Phi,
\end{equation}
where
\begin{align}
       && \left(\frac{\partial G}{\partial T}\right)_{\Omega_a, \Omega_b, \Phi} = -S,\hspace{0.5 in}  
        \left(\frac{\partial G}{\partial \Omega_{a}}\right)_{T, \Phi, \Omega_b} = -J_{a},  \nonumber\\
       && \left(\frac{\partial G}{\partial \Omega_{b}}\right)_{T, \Phi, \Omega_b} = -J_{b}, \hspace{0.5 in}  
        \left(\frac{\partial G}{\partial \Phi}\right)_{T, \Omega_a, \Omega_b} = -Q.
\end{align}

\par 

Notice that the above quantum-statistical relation which we call the Gibbs-Duhem relation is going to be the same for the background and counterterms methods apart from a modification in mass and action calculations. As we have mentioned in the Introduction, in the counterterms method, this relation plays an important role to prove the validity of the first law.





{\bf A Generalized Smarr's Formula}


For asymptotically AdS spaces, in order to construct a generalized Smarr's formula, one must allow the cosmological constant to vary, leading to the following form of Smarr's formula \cite{pope2011} 
\begin{align} \label{Smarr}
    M_0 = \frac{3}{2} \left(TS + \Omega_a J_a + \Omega_b J_b \right) + \Phi Q - PV_0,
\end{align}

where the cosmological constant plays the role of a pressure with some thermodynamic volume as a conjugate variable \cite{Kastor:2009wy,Kubiznak:2012wp}, or 
\begin{align} \label{P}
    P = - \frac{\Lambda}{8\pi} =\frac{3g^2}{4\pi},
\end{align}
And $V_0$ is the ``thermodynamic volume" $V_0$ conjugate to $P$. Following the prescription in \cite{pope2011}, we find
\begin{align}
    V_0 &= \left(\frac{\partial M_0}{\partial P}\right)_{S, Q, J_a, J_b} \nonumber \\
    &=\frac{\pi^{2}}{2 \Xi_{a} \Xi_{b}}\left[\left(r_{+}^{2}+a^{2}\right)\left(r_{+}^{2}+b^{2}\right)+\frac{2}{3} a b q\right]+\frac{2 \pi}{3}\left(a J_{a}+b J_{b}\right)  \label{V_0} ,
\end{align}
in accordance with \cite{grunau}. With some algebraic manipulations, it is straightforward to show that the above Smarr's formula \eqref{Smarr} is satisfied.



%% file: 4-Counterterms_and_the_First_Law.tex
\section{Thermodynamics: Counterterms Method} \label{sec:CTs}

 \label{sec:CTs_and_the_first_law}

As mentioned in the introduction the need to use the counterterms method is to obtain a finite on-shell gravitational action. This method gives a non-ambiguous procedure to regularise the action in a manner independent of any other spacetime (i.e. a background spacetime). This allows us to make some important connections between the bulk gravity and the dual CFT in four dimensions. As we will see in the coming sections, this method enables us to calculate quantities such as the vacuum expectation value (vev) of the stress-tensor, the conformal anomaly, as well as the Casimir energy of the boundary field theory. Our aim is to calculate the gravitational finite action first, then use it to calculate the vev of the field theory stress tensor through calculating the quasilocal stress tensor, or the Brown-York stress-tensor \cite{brown_york} (BY) (using eq. (\ref{h/gamma_factor}). For details see for example\cite{kraus,awad_johnsson1}). The BY tensor is also used to calculate conserved quantities such as the mass and angular momenta.

\subsection{Action Calculation}

The action of the Lagrangian presented in eq.(\ref{pope_lagrangian}) is composed of the following terms
\begin{equation}
    I_\text{non-ren} =  I_\text{EH} + I_\text{GH} + I_\text{EM}.
\end{equation} 
Calculating the electromagnetic part\footnote{See also, \cite{silva2006}}, one finds

\begin{equation} \label{I_EM}
    I_\text{EM} = \frac{-\pi \beta}{4 \Xi_a \Xi_b} \left[\frac{q^{2} r_{+}^{2}}{\left(r_{+}^{2}+a^{2}\right)\left(r_{+}^{2}+b^{2}\right)+a b q}\right], 
\end{equation}

While the Einstein-Hilbert action calculation leads to 
\begin{equation} \label{Ieh_1}
    I_\text{EH} = \beta \left[ \frac{\pi  g^2 r^4}{4 \Xi _a \Xi _b} + \frac{\pi  g^2 r^2 \left(a^2+b^2\right)}{4 \Xi _a \Xi _b} - \frac{\pi  g^2 r_+^2 \left(a^2+b^2+r_+^2\right)}{4 \Xi_a \Xi_b} \right] \Bigg|_{r\to+\infty}.
\end{equation}

And the Gibbons-Hawking action calculation produces
\begin{align} \label{Igh_1}
    I_\text{GH} &= \beta \bigg[-\frac{\pi  g^2 r^4}{\Xi _a \Xi _b} + \frac{15 \pi  r^2 \left(\Xi_a +\Xi_b + 3/4\right)}{24 \Xi _a \Xi _b}     +\frac{\pi  \left(a^4 g^2-8 a^2 b^2 g^2-9 a^2+b^4 g^2-9 b^2+24 m\right)}{24 \Xi _a \Xi _b} \bigg]\bigg|_{r\to+\infty}.
\end{align}

Summing the previous three expressions we get the following divergent action
\begin{align} \label{I_non-reg1}
    \nonumber I_\text{non-ren} =  & I_\text{EH} + I_\text{GH} + I_\text{EM}\\
    = &\left[-\frac{3 \pi \beta g^2 r^4}{4 \Xi _a \Xi _b} + \beta r^2 \left(\frac{\pi  g^2 \left(a^2+b^2\right)}{4 \Xi _a \Xi _b}+\frac{\pi  \left(-15 a^2 g^2-15 b^2 g^2-18\right)}{24 \Xi _a \Xi _b}\right)\right]\Bigg|_{r\to+\infty} + \text{finite terms}.
\end{align}

Now let us calculate the counterterms and verify that they cancel these divergences. The expression for these counterterms is given in \cite{surface_terms} by
\begin{align} \label{I_ct}
    I_{\mathrm{ct}}=\frac{1}{8 \pi} \int_{\partial \mathcal{M}} \text{d}^{n} x \sqrt{h}\left[\frac{n-1}{\ell}+\frac{\ell}{2(n-2)} \mathcal{R}+\frac{\ell^{3}}{2(n-4)(n-2)^{2}}\left(\mathcal{R}_{a b} \mathcal{R}^{a b}-\frac{n}{4(n-1)} \mathcal{R}^{2}\right)+\ldots\right],
\end{align}

where $\mathcal{R}_{ab}$ and $\mathcal{R}$ are the Ricci tensor and Ricci scalar of the boundary metric, respectively, and we recall that $\ell=1/g$. These counterterms were calculated in $D=n+1$ dimensions \cite{kraus,surface_terms}. The extra terms denoted by ``$...$" are needed in higher dimensions. Evaluation of the counterterms in \eqref{I_ct} yields

\begin{equation} \label{I_ct2}
     I_\text{ct} = \beta \left[ \frac{3 \pi  g^2 r^4}{4 \Xi _a \Xi _b} + \frac{3 \pi  r^2 \left(a^2 g^2+b^2 g^2+2\right)}{8 \Xi _a \Xi _b} \right]\Bigg|_{r\to+\infty} + \text{finite terms}.
\end{equation}

The final expression for the Euclidean action from the counterterms subtraction method is found by adding the term in \eqref{I_ct2} to the divergent action in \eqref{I_non-reg1}, which gives the regulated action
\begin{align} \label{I_ren_Hassan}
    \begin{split}
        I_\text{ren} = &\frac{\pi \beta}{96 g^2 \Xi_a \Xi_b} \big[ \Xi_a^2 + \Xi_b^2 + 7 \Xi_a  \Xi_b + 24 g^2 \big( m - a^2b^2g^2 - g^2r_+^2 (r_+^2 + a^2 + b^2) \big) \big] \\
        &- \frac{\pi \beta}{4 \Xi_a \Xi_b} \left[\frac{q^{2} r_{+}^{2}}{\left(r_{+}^{2}+a^{2}\right)\left(r_{+}^{2}+b^{2}\right)+a b q}\right].
    \end{split}
\end{align}

This action reduces to that of the Reissner-Nordström black hole in \cite{catastrophic_holography} in the limit $a,b \to 0$. We can also re-express the gravitational part of the action as 
\begin{align} \label{I_grav}
    \nonumber I_\text{ren}^\text{grav} = \beta M_\text{BG}^{\text{Kerr-AdS}_5} + \frac{\pi \beta \left[m - g^2(r_+^2+a^2)(r_+^2+b^2) \right]}{4\Xi_a \Xi_b},
\end{align}

where 

\begin{equation} \label{Mc_Kerr-AdS5}
    M_\text{BG}^{\text{Kerr-AdS}_5} = \frac{\pi \left[9 \Xi_a 9\Xi_b + (\Xi_a - \Xi_b)^2\right]}{96g^2\Xi_a\Xi_b}
\end{equation}

is the background energy of the Kerr-AdS$_5$ black hole \cite{skenderis}. Our result takes the exact same \textit{form} as the one calculated by Papadimitriou and Skenderis \cite{skenderis} for a non-charged rotating black hole. 

The difference here of course is that the charge does make an appearance in the $\beta$ term. Nevertheless, it is interesting that the gravitational action maintains the same expression in terms of $\beta$. This also automatically shows that our result reduces to that in \cite{skenderis} since, in the absence of an electric charge, the temperature takes the same form as that in \cite{skenderis}, and the electromagnetic action vanishes.

\subsection{Mass and Angular Momenta Calculations} \label{sec:calculation_of_M_and_J}

The calculation of conserved charges is based on the Brown-York quasilocal charge definition \cite{brown_york}. The Brown-York quasilocal stress tensor is given by
\begin{align} \label{brown_york_stress_tensor}
    T^{a b} &= \frac{-2}{\sqrt{|h|}} \frac{\delta I_\text{ren}}{\delta h_{ab}}\nonumber \\
    &= \frac{1}{8 \pi G}\left[K^{a b}-h^{a b} K+ g(n-1)h^{a b} -\frac{\mathcal{G}^{a b}}{g(n-2)}\right],
\end{align}
where $h_{ab}$, $K_{ab}$, and $\mathcal{G}_{ab}$ are the metric, extrinsic curvature, and Einstein tensors on the induced boundary, respectively. Notice that $D=n+1$ is the spacetime dimensions. A charge associated with a Killing vector $K^a$ is defined by

\begin{equation} \label{BY_quasilocal_charge}
    Q[K] = \int_{S^3} \text{d}^3x \sqrt{\sigma} u^a T_{ab} K^b.
\end{equation}

Here $u_a = -N \nabla_a t$, while $N$ is the lapse function and $\sigma$ is the spacelike metric that appear in the ADM-like decomposition of the metric

\begin{equation} \label{metric_foliation}
    ds^2 = -N \text{d}t^2 + \sigma_{ab}(\text{d}x^a + N^a \text{d}t)(\text{d}x^b + N^b \text{d}t).
\end{equation}

In the above expression $N^a$ is the shift vector. The Killing vector associated with the mass is the timelike Killing vector $\chi = \partial_t$. Plugging this in \eqref{BY_quasilocal_charge} initially yields a slightly complicated expression for the mass. To simplify the long calculation let us write the mass as the sum of four terms:

\begin{equation}
    M = M_\text{nr1} + M_\text{nr2} + M_\text{ct1} + M_\text{ct2}.
\end{equation}

The first two terms are divergent; they result from the original non-regularized components of the Brown-York quasilocal stress tensor. The last two terms are those arising from the counterterms action. The first of those four terms is given by
\begin{align}
    \nonumber M_\text{nr1} = &\frac{\pi  m \left[-3 g^2 \left(a^2+b^2\right)+a^2 g^2 \Xi _b+b^2 g^2 \Xi _a-2 \Xi _a \Xi _b+6\right]}{8 G \Xi _a^2 \Xi _b^2} \\
    \nonumber &+ \frac{\pi  q \left[a b g^2 \left(-3 g^2 \left(a^2+b^2\right)+a^2 g^2 \Xi _b+6\right)+a b^3 g^4 \Xi _a+2 a b g^2 \Xi _a \Xi _b\right]}{8 G \Xi _a^2 \Xi _b^2} \\
    &+ \frac{\pi  \left[2 a^2 b^2 g^2 \Xi _a \Xi _b+2 g^2 r^2 (a^2+b^2) \Xi _a \Xi _b+2 g^2 r^4 \Xi _a \Xi _b\right]}{8 G \Xi _a^2 \Xi _b^2},
\end{align}

where we have reinstated the gravitational constant $G$ (originally a factor in the denominator of the Brown-York quasilocal stress tensor) for future need. The second term is given by
\begin{align}
    M_\text{nr2} = & \frac{-\pi}{24 G \Xi _a \Xi _b} \Big[a^4 -g^2 + 3 r^2 \big(5 g^2 \left(a^2+b^2\right)+6\big)+8 a^2 b^2 g^2+9 a^2-b^4 g^2+9 b^2     &+24 g^2 r^4 -24 m\Big].
\end{align}

The first part of the counterterms contribution is
\begin{align}
    \nonumber M_\text{ct1} = &\frac{\pi }{32 g^2 G \Xi _a^2 \Xi _b^2} \Big[ 12 g^2 \left(a^2 g^2 \Xi _b+a^2 g^2+b^2 g^2-2\right) \left(a b g^2 q+m\right)+\Xi_a \Big(\Xi _b \big(-a^4 g^4+a^2 g^2 \\
    \nonumber &\quad \left(11 b^2 g^2+18 g^2 r^2+9\right)+24 a b g^4 q-b^4 g^4+9 b^2 \left(2 g^4 r^2+g^2\right)+3 \big(8 g^4 r^4+4 g^2 r^2\\
    &\quad -1\big)\big)+12 b^2 g^4 \left(a b g^2 q+m\right)\Big) \Big].
\end{align}

And lastly, the second contribution from the counterterms gives

\begin{equation}
    M_\text{ct2} = -\frac{3 \pi  \left[a^2 g^4 \left(b^2+r^2\right)+g^2 r^2 \left(b^2 g^2-2\right)-1\right]}{16 g^2 G \Xi_a \Xi_b}.
\end{equation}

The addition of all these terms gives the mass via the counterterms subtraction method,
\begin{align} \label{mass_Hassan_ct}
    M = \frac{\pi  m \left(2 \Xi _a + 2 \Xi _b - \Xi _a\Xi _b\right) + 2 \pi  a b g^2 q \left(\Xi _a+\Xi _b\right) }{4 G \Xi _a^2 \Xi _b^2} + \frac{\pi \left[ 9\Xi_a\Xi_b + (\Xi_a - \Xi_b)^2 \right]}{96G g^2 \Xi_a \Xi_b}.
\end{align}

The first term on the RHS is just the black hole mass calculated in \cite{pope} by integrating the first law. The second term is the background energy of the spacetime: it is the value that the total energy reduces to in the absence of the black hole (when $m=0$, $q=0$). We note that this background energy is the same as that in the Kerr-AdS$_5$ black hole solution \cite{awad_johnsson3}. In $\S$\ref{sec:casimir_energy} we will show that this is exactly equal to the Casimir energy of the dual CFT on the boundary. Furthermore, when the two rotation parameters are set to zero, this vacuum energy in \eqref{M_BG} reduces to the background energy of the pure non-rotating AdS$_5$ spacetime given in \cite{kraus} by

\begin{equation}
    M_\text{BG}\Big|_{a=b=0} = \frac{3 \pi}{32g^2 G}.
\end{equation}

\par 

The full expression for the mass calculated via the counterterms method can now be written as
\begin{align} \label{M=M0+Mct}
    M = M_0 + M_\text{BG},
\end{align}

with 
\begin{equation} \label{M_BG}
    M_\text{BG} = \frac{\pi \left[ 9\Xi_a\Xi_b + (\Xi_a - \Xi_b)^2 \right]}{96G g^2 \Xi_a \Xi_b}.
\end{equation}

\par

The final quantities to calculate using the counterterms method are the angular momenta. To get $J_a$ (respectively $J_b$) we plug in \eqref{BY_quasilocal_charge} the Killing vector $\partial_\phi$ (respectively $\partial_\psi$). The angular momentum in the $\phi$-direction is given by
\begin{align} \label{J_a_Hassan}
    J_a &= \int_{S^3} \frac{\sin^3\theta \cos\theta \left[ am + b\left( a^2g^2q + \frac{1}{2}q\Xi_a \right) \right]}{2\pi \Xi_a^2 \Xi_b} \text{d}\theta \text{d}\phi \text{d}\psi \nonumber \\ 
    &= \frac{\pi\left[2 a m+q b\left(1+a^{2} g^{2}\right)\right]}{4 \Xi_{a}^{2} \Xi_{b}}.
\end{align}



And we find a similar result for $J_b$,
\begin{align} \label{J_b_Hassan}
    J_b &= \frac{\pi\left[2 b m+q a\left(1+b^{2} g^{2}\right)\right]}{4 \Xi_{a} \Xi_{b}^{2}}.
\end{align}

These results are equal to the angular momentum calculated in \cite{pope} using the Komar integral.





There are some intriguing features of this solution worth discussing, some of which are indeed unique. Before we discuss these features let us take some limits to check the above mass/total energy expression. By setting $q=0$ and $a=b=0$, we reduce this solution to the Schwarzschild mass, 

\begin{equation*}
    M\big|_{q=0,a=b=0} = \frac{3 \pi m}{4}.
\end{equation*}

This means that the parameter $m$ has the usual interpretation of a ``mass parameter". Yet there seems to be an unusual aspect of this solution: when this parameter vanishes, the black hole's mass does not vanish not only because of the background energy \textit{but also due to contributions from the electric charge, $q$.}

\begin{equation}
    M\big|_{m=0} = \frac{\pi abg^2q(\Xi_a+\Xi_b)}{2\Xi_a^2\Xi_b^2} + M_\text{BG}.
\end{equation}

Indeed, the charge contributes to the total mass of the solution! We do not know any other solution that shares this property with the solution presented here. For instance, we find the mass of a static charged black hole in five-dimensional anti-de Sitter spacetime and the neutral Kerr-AdS solutions have no charge contributions in their mass expressions. Therefore, the physical significance of the expression in \eqref{mass_Hassan_ct} when $m$ vanishes seems a bit puzzling. However, one can show that there are no horizons, or real values for $r_+$, if $m=0$ while $q\neq0$. In other words, the case $m=0, q\neq0$ has a naked singularity.

\par 

Another puzzling feature appears in the angular momenta expression in equations (\ref{J_a_Hassan}), and (\ref{J_b_Hassan}). The angular momenta carry some dependence on the electric charge. This feature has no analog in four dimensions. While the mass is non-vanishing if $m$ goes to zero, there are also non-vanishing angular momenta $J_a$ and $J_b$. Furthermore, if we keep $m \neq 0$ and set $a = 0$ we still have a non-null value for $J_a$! The same thing happens if we $m \neq 0$ and $b = 0$: we still have a non-null value for $J_b$! Of course, this does not happen in the Kerr-AdS$_5$ solution where the angular momenta are given \cite{skenderis} by
\begin{align} \label{skenderis_angular_momenta}
    J_{a} &=\frac{\pi a m}{2 \Xi_{a}^{2} \Xi_{b}}, \quad J_{b}=\frac{\pi b m}{2 \Xi_{b}^{2} \Xi_{a}}.
\end{align}
One is then left to wonder if these new features can produce some interesting phenomena for these black holes.

%% file: 5-Holography.tex
\section{Holographic Stress Tensor and Anomaly} \label{sec:holography}

One of the predictions of the AdS/CFT correspondence is that the conformal anomaly calculated from the conformal field theory on the boundary should exactly match the trace of the gravitational stress tensor (BY tensor) \cite{kraus,awad_johnsson1}, where the identification $G^{-1} \leftrightarrow 2N^2g^3/\pi$ is to be made. Another prediction is the matching of the dual field theory Casimir energy with the background energy obtained as we set $m=0$ and $q=0$ in the mass expression. In this section we will aim to verify these predictions as well as calculate the vev of the renormalized CFT stress tensor predicted by the duality. 

\par 

\subsection{CFT Stress Tensor and Conformal Anomaly} \label{sec:conformal_anomaly}

It is important to notice here that the metric in eq.(\ref{metric}) is asymptotically AdS with a conformal boundary (with $\mathbb{R} \times S^{3}$ topology). The induced metric at the boundary is given by 
\begin{equation} \label{boundary_metric}
    ds_\text{Boundary}^2 = g^2 r^2 \left[ -\text{d}t^2 + \frac{2a \sin^2\theta}{\Xi_a} \text{d}t \text{d}\phi + \frac{2b \cos^2\theta}{\Xi_b} \text{d}t \text{d}\psi + \frac{\text{d}\theta^2}{g^2 \Delta_\theta(\theta)} + \frac{\sin^2\theta}{g^2 \Xi_a}\text{d}\phi^2 + \frac{\cos^2\theta}{g^2 \Xi_b}\text{d}\psi^2 \right],
\end{equation}

which is nothing but the rotating Einstein Universe. 
The CFT or the boundary metric  is found by removing the divergent conformal factor $g^2 r^2$ from the above metric \cite{awad_johnsson1}

\begin{equation} \label{cf_metric}
    ds_\text{BG}^2 = -\text{d}t^2 + \frac{2a \sin^2\theta}{\Xi_a} \text{d}t \text{d}\phi + \frac{2b \cos^2\theta}{\Xi_b} \text{d}t \text{d}\psi + \frac{\text{d}\theta^2}{g^2 \Delta_\theta(\theta)} + \frac{\sin^2\theta}{g^2 \Xi_a}\text{d}\phi^2 + \frac{\cos^2\theta}{g^2 \Xi_b}\text{d}\psi^2.
\end{equation}

The bulk stress tensor is related to the expectation value of the renormalized CFT stress tensor $\langle \hat{T}_{ab} \rangle$ \cite{awad_johnsson1} by

\begin{equation} \label{h/gamma_factor}
    \sqrt{-\gamma}\, \gamma_{a b} \langle \hat{T}^{b c} \rangle =\lim _{r \rightarrow +\infty} \sqrt{-h}\, h_{a b} T^{b c}.
\end{equation}

Therefore, we expect the trace of the gravitational tensor to be related to the CFT's stress tensor through the following factor 
\begin{equation}
    \lim_{r \to +\infty} \sqrt{h/\gamma} = g^4 r^4.
\end{equation}

Evaluation of the trace of \eqref{brown_york_stress_tensor} gives

\begin{equation}
    T^a {}_a = -\frac{\left(a^2-b^2\right) \Big[3 g^2 \left(a^2-b^2\right) \cos^4\theta-2 \cos^2\theta \left(a^2 g^2-2 b^2 g^2+1\right)-b^2 g^2+1\Big]}{8 \pi  g G r^4}.
\end{equation}

Multiplying this by the conformal factor $g^4 r^4$ yields

\begin{equation} \begin{aligned} \label{conformal_anomaly_Hassan_gravity}
    g^4 r^4 T^a {}_a =  -\frac{g^3 \left(a^2-b^2\right) \Big[3 g^2 \left(a^2-b^2\right) \cos^4\theta-2 \cos^2\theta \left(a^2 g^2-2 b^2 g^2+1\right)-b^2 g^2+1\Big]}{8 \pi G}.
\end{aligned} \end{equation}

This result matches that found in \cite{awad_johnsson3} for the Kerr-AdS$_5$ solution. This is expected since the difference between the Kerr-AdS$_5$ metric and that in \eqref{metric} is a term proportional to the charge parameter $q$, which vanishes at the boundary.

\par 

We now look at the expectation value of the stress tensor of the dual CFT. The details that we followed in calculating and renormalizing this tensor are outlined in Chapter 6 of ref. \cite{birrel_davies}. Following the notation in ref. \cite{awad_johnsson3}, the renormalized stress tensor is given by

\begin{equation}
    \langle \hat{T}_{a b} \rangle = -\frac{1}{16\pi^2} \sum_s\left(\frac{1}{9} \alpha^s H_{ab}^{(1)} + \beta^s H_{ab}^{(3)} \right),
\end{equation}

where the summation is over the possible fields of the theory, with $s=0,\frac{1}{2},1$ standing for scalar, Weyl spinor, and U(1) gauge fields, respectively. If we calculate the expectation value over a vacuum state then $H_{ab}^{(4)}$ will vanish since it is the vacuum expectation value of the stress tensor in flat spacetime. 

\par 

Furthermore, one can choose a regularization scheme in which the $\alpha^s$ coefficients vanish. For more details on this, the reader can refer to ref. \cite{awad2016}. The values of the remaining $\beta^s$ coefficients are given \cite{awad2016} by 
\begin{align}
    & \beta^0 = -\frac{1}{2880\pi^2} N^0, \quad \beta^\frac{1}{2} = -\frac{1}{2880\pi^2} N^\frac{1}{2}, \quad \beta^1 = -\frac{1}{2880\pi^2} N^1,
\end{align}
where $N^i$ is the number of fields of spin $i$. These numbers are \cite{awad_johnsson1} $N^0=6N^2$, $N^\frac{1}{2}=4N^2$ and $N^1=N^2$. The tensor $H_{ab}^{(3)}$ is given in \cite{birrel_davies} by

\begin{equation}
    H_{a b}^{(3)}=\frac{1}{12} \mathcal{R}^{2} \gamma_{a b}-\mathcal{R}^{c d} \mathcal{R}_{c a d b},
\end{equation}

where $\gamma_{ab}$ is the CFT metric tensor and $\mathcal{R}_{abcd}$, $\mathcal{R}_{ab}$ and $\mathcal{R}$ are the Riemann tensor, Ricci tensor, and Ricci scalar of the CFT, respectively. 

\par

We list below the non-vanishing components of the CFT stress tensor:
\begin{align}
    \langle \hat{T}_{tt} \rangle = &-\frac{495N^2  g^4}{1952 \pi^2 \Xi _a \Xi _b}  \Big[ \Big(g^2 \left(a^2+b^2\right)+g^2 (a^2-b^2) \cos (2 \theta )-2\Big) \Big(g^2 (a^2-b^2) \big(20 \cos (2 \theta ) \left(g^2 \left(a^2+b^2\right)-2\right) \nonumber\\
        &+7 g^2 (a^2-b^2) \cos (4 \theta )\big)-24 g^2 \left(a^2+b^2\right)+g^4 \left(5 a^4+14 a^2 b^2+5 b^4\right)+24\Big) \Big],\nonumber\\
    \langle \hat{T}_{\theta \theta} \rangle= &\frac{495 N^2 g^2 }{976 \pi^2 \left(g^2 \left(b^2-a^2\right) \cos ^2\theta -b^2 g^2+1\right)}  \Big[ g^2 \left(a^2-b^2\right) \Big(3 g^2 \left(a^2-b^2\right) \cos (4 \theta )+4 \cos (2 \theta ) \big(g^2 \left(a^2+b^2\right)\nonumber\\
        &-2\big)\Big)+8 g^2 \left(a^2+b^2\right)+g^4 \big(-7 a^4+6 a^2 b^2-7 b^4\big)-8 \Big],\nonumber\\
    \langle \hat{T}_{\phi \phi} \rangle = &\frac{495 N^2 g^2 \sin ^2\theta }{976 \pi^2 \Xi _a}  \Big[ g^2 (a^2-b^2) \Big(7 g^2 (a^2-b^2) \cos (4 \theta )-4 \cos (2 \theta ) \big(g^2 \left(3 a^2-5 b^2\right)+2\big)\Big)+8 g^2 \big(a^2+b^2\big)\nonumber\\
        &+g^4 \big(5 a^4-18 a^2 b^2+5 b^4\big)-8 \Big],\nonumber\\
    \langle \hat{T}_{\psi \psi} \rangle = &\frac{495 N^2 g^2 \cos ^2\theta }{976 \pi^2 \Xi _b}  \Big[ g^2 \left(a^2-b^2\right) \Big(7 g^2 \left(a^2-b^2\right) \cos (4 \theta )+4 \cos (2 \theta ) \big(g^2 \left(5 a^2-3 b^2\right)-2\big)\Big)+8 g^2 \big(a^2+b^2\big)\nonumber\\
        &+g^4 \big(5 a^4-18 a^2 b^2+5 b^4\big)-8 \Big].
\end{align}

\par

It is important to state here that the above CFT stress tensor matches the one predicted by the duality, which is calculated using Brown-York tensor and the above relation in eq. (\ref{h/gamma_factor}). Indeed, this is one of the important non-perturbative silent checks of this duality. As a result, evaluating the trace of this stress tensor yields the expected conformal anomaly
\begin{align} \label{cft_anomaly_Hassan}
   \langle \hat{T}^{a} {}_{a} \rangle = 
   -\frac{\left(a^{2}-b^{2}\right) N^{2} g^6 \left[  3 g^2 \left(a^2-b^2\right) \cos^4\theta-2 \cos^2\theta \left(a^2 g^2-2 b^2 g^2+1\right)-b^2 g^2+1\right]}{4 \pi^{2} }.
\end{align}

This identification is exact upon remembering that, in the AdS/CFT correspondence, $2N^2g^3/\pi$ is translated to $G^{-1}$ at the gravity side and vice versa. We have therefore shown that the CFT stress tensor -- which is calculated from the counterterm-regulated action at the gravity side -- is exactly equal to the CFT boundary stress tensor calculated at the field theory side. This also implies an equality between the conformal anomalies on both sides.

\subsection{Casimir Energy} \label{sec:casimir_energy}

Another consequence of the matching of the above two stress tensors -- the one predicted by gravity and that calculated on the field theory side -- is the matching between the vacuum energies, \eqref{M_BG} on the gravity side and the Casimir energy $E_\text{Casimir}$ on the field theory side. The Casimir energy is found using the formula (\cite{awad_johnsson1})

\begin{equation} \label{Casimir_en_expression}
    E_\text{Casimir}=\sum_{s=0, \frac{1}{2}, 1} N^{s} \int_{S^3} \mathrm{d}^{3} x \sqrt{\sigma}\, \chi^{a} \langle \hat{T}_{a b}^{s} \rangle u^{b}.
\end{equation}

Here the summation is again over the possible fields of the theory. $\chi^a$ and $u^a$ are the timelike Killing vector and unit normal vector, and $\sigma_{ab}$ is the foliation metric of the conformal boundary. The conformal foliation metric is found using

\begin{equation}
    \sigma_{ab} = g^2 r^2 \left(g_{ab} + u_a u_b \right),
\end{equation}

whose determinant is given by 

\begin{equation}
    \sigma = \frac{\left(a^{2} \cos^{2}\theta + b^{2}\sin^2\theta +r^{2}\right) \sin^{2}\theta \cos^{2}\theta}{g^{6} r^{2} \Delta_{\theta} \Xi_{a} \Xi_{b}}.
\end{equation}

Direct evaluation of the integral in \eqref{Casimir_en_expression} followed by some simplifications gives

\begin{equation} \label{casimir_en_Hassan}
    E_\text{Casimir} =  \frac{N^2 g \left[  9 \Xi_a \Xi_b + (\Xi_a - \Xi_b)^2 \right]}{48 \Xi_a \Xi_b}.
\end{equation}

Making the identification $\pi/(2Gg^3) \leftrightarrow N^2$, it is easy to see that the Casimir energy in \eqref{casimir_en_Hassan} is identical to the background energy of the bulk spacetime in \eqref{M_BG},
\begin{equation} \begin{aligned}
    E_\text{Casimir} = M_\text{BG}.
\end{aligned} \end{equation}

%% file: 6-Thermodynamics_with_M.tex
\section{The First Law and the Counterterms Method} \label{sec:thermodynamics_with_CTs}

As was discussed in \cite{skenderis}, introducing surface counterterms to regulate the on-shell gravitational action in five dimensions induces an anomalous Weyl transformation on the boundary.\footnote{Notice that this happens when the rotation parameters are different.} This is how the bulk encodes the boundary Weyl anomaly. Let us denote the radial spacelike boundary at infinity by $\partial \mathcal{M}$. It was found \cite{skenderis} that a Weyl factor $\delta \sigma$ changes the renormalized action by

\begin{equation} \label{first_law_skenderis}
    \delta_\sigma I_\text{ren} = \int_{\partial \mathcal{M}} \text{d}^nx \sqrt{-h} \mathcal{A} \delta\sigma,
\end{equation}

in the presence of a conformal anomaly $\mathcal{A}$.

\par

Now, we have seen that in order to verify the first law, we need to vary various quantities, such as the entropy, angular momenta, and charge with respect to $r_+$, $q$, $a$ and $b$ (see for example eq. \eqref{first_law_3_alpha}). But since the boundary metric is expressed in terms of $a$ and $b$, variation of these parameters will change the boundary metric that should remain fixed. We will see (eq. \eqref{variation_of_bg_metric}) that variations of $a$ and $b$ have the effect of a Weyl transformation on the boundary metric. The entropy, angular momenta, and electric charge are expressed as surface integrals of fluxes whose sources are localized in the black hole region. In that sense, they can be written as integrals over the horizon \cite{skenderis}. Thus they will not be affected by variations of the boundary metric. This is not the case for the mass. As we saw in eq. \eqref{mass_Hassan_ct}, the total mass is the sum of a term that depends on the mass parameter ($M_0$) plus the background energy ($M_\text{BG}$). The latter is certainly affected by conformal transformations of the boundary. Another quantity that is affected by the boundary conformal transformation is the renormalized action, which contains contributions from surface integrals on the boundary. 

\par 

The main idea is to subtract from the LHS of the first law the variation of the mass that results from varying the boundary metric. If we denote this mass variation by $\delta_\sigma M$, the correct form of the first law \cite{skenderis} is thus given by
\begin{align} \label{expected_first_law}
    d M &= \delta_\sigma M + T d S + \Omega_a d J_a + \Omega_b d J_b + \Phi d Q.
\end{align}
Note that the variations of the renormalized action $I_\text{ren}$ and mass are not independent since (see eq. \eqref{quantum_statistical_relation} below)

\begin{equation}
    I_\text{ren} = \beta (M - T S-\Omega_a J_a-\Omega_b J_b-\Phi Q),
\end{equation}

so

\begin{equation}
    \delta_\sigma I_\text{ren} = \beta \delta_\sigma M.
\end{equation}

We argue here that, by a similar analysis, the volume term (see eq. \eqref{V_0}) needs to be modified due to extra terms arising from boundary variation when the AdS radius $1/g$ is varied as well.

\par 

In conclusion, we expect the action and mass calculated from the counterterms method to verify the quantum statistical relation, the first law, and Smarr's formula. But in order to show the latter two, we must account for extra terms which result from a Weyl transformation of the boundary when certain parameters are varied

\subsection{The First Law in the Presence of a Conformal Anomaly} \label{sec:verification_of_first_law}

We begin by noting that the counterterms action \eqref{I_ren_Hassan} and mass \eqref{mass_Hassan_ct} satisfy the so-called quantum statistical relation

\begin{equation} \label{quantum_statistical_relation}
    I_\text{ren} = \beta G(T, \Omega_a, \Omega_b, \Phi).
\end{equation}

We will adopt the same procedure that was done in \cite{skenderis}. The first step is to transform the boundary metric into a more canonical asymptotic form. To do so we use the coordinates $\bar{r}$ and $\bar{\theta}$ defined by
\begin{align}
    r &=\bar{r}\left[1+\frac{\hat{\Delta}_{\bar{\theta}}}{4g^2 \bar{r}^2} +\frac{\hat{\Delta}_{\bar{\theta}}}{16 g \bar{r}} \left(1+\hat{\Xi}_{a}+\hat{\Xi}_{b}-2 \hat{\Delta}_{\bar{\theta}}\right) +\mathcal{O}\left(\frac{1}{\bar{r}^{6}}\right)\right],\\
    \theta &= \bar{\theta}+\frac{1}{16 g \bar{r}}\left(1-\hat{\Delta}_{\bar{\theta}}\right) \hat{\Delta}_{\bar{\theta}}^{\prime} -\frac{1}{32 g^6 \bar{r}^6}\left(1-\hat{\Delta}_{\bar{\theta}}\right) \hat{\Delta}_{\bar{\theta}}^{\prime}\left(1+\hat{\Xi}_{a}+\hat{\Xi}_{b}+3 \hat{\Delta}_{\bar{\theta}}\right) +\mathcal{O}\left(\frac{1}{\bar{r}^{8}}\right).
\end{align}

The functions $\hat{\Delta}_\theta$, $\hat{\Xi}_a$ and $\hat{\Xi}_b$ are given by
\begin{align}
    \hat{\Delta}_{\theta} &=1-\Delta_{\theta},\quad \hat{\Xi}_{a} =1-\Xi_{a},\quad \hat{\Xi}_{b} = 1-\Xi_{b}.
\end{align}

The conformal boundary metric in terms of the new coordinates is 
\begin{equation} \label{bg_metric_bar_coordinates}
    d \bar{s}^{2}_\text{BG}= -\text{d} t^{2}+\frac{2 a \sin ^{2} \bar{\theta}}{\Xi_{a}} \text{d} t \text{d} \phi+\frac{2 b \cos ^{2} \bar{\theta}}{\Xi_{b}} \text{d} t \text{d} \psi+\frac{1}{g^2 \Delta_{ \bar{\theta}}} \text{d} \bar{\theta}^{2}+\frac{ \sin ^{2} \bar{\theta}}{g^2 \Xi_{a}} \text{d} \phi^{2}+\frac{\cos ^{2} \bar{\theta}}{g^2 \Xi_{b}} \text{d} \psi^{2}.
\end{equation}


We now consider the variation of this metric under infinitesimal variations of $a$ and $b$. We switch to a new coordinate system $(t, \bar{r}, \bar{\theta}^\prime, \phi^\prime, \psi^\prime)$ given by
\begin{equation}
    \tan^{2} \bar{\theta}=\left(1+\frac{\delta \Xi_{a}}{\Xi_{a}}-\frac{\delta \Xi_{b}}{\Xi_{b}}\right) \tan ^{2} \bar{\theta}^{\prime}, \quad \phi=\phi^{\prime}-g^2 t\delta a, \quad \psi=\psi^{\prime}-g^2 t\delta b.
\end{equation}

In terms of these coordinates, the resulting variation of the metric in \eqref{bg_metric_bar_coordinates} is given \cite{skenderis} by

\begin{equation} \label{variation_of_bg_metric}
    d \bar{s}_\text{BG}^{2} \rightarrow\left(1-\frac{\delta \Xi_{a}}{\Xi_{a}} \sin ^{2} \bar{\theta}-\frac{\delta \Xi_{b}}{\Xi_{b}} \cos ^{2} \bar{\theta}\right) d \bar{s}_\text{BG}^{2}.
\end{equation}

We use the value of the conformal anomaly that we calculated in \eqref{cft_anomaly_Hassan} to find the variation of the action resulting from the Weyl factor in \eqref{variation_of_bg_metric}:
\begin{align} 
    \delta_\sigma I_\text{ren} &=  \int_{ [0, \,\beta] \times S^{3}} \text{d}^4x \sqrt{-h} \mathcal{A} \delta\sigma \label{delta_I_in_A} \\
    &= \frac{\pi \beta}{96 g^2 G} \delta \left(\frac{\Xi_{a}}{\Xi_{b}}+\frac{\Xi_{b}}{\Xi_{a}}\right). \label{delta_I}
\end{align}

This form is in agreement with \cite{skenderis}, again with the caveat that the electric charge appears in $\beta$. We also explicitly verify that 
\begin{align} \label{delta_I=beta_delta_M}
    \delta_\sigma I_\text{ren} &= \beta \delta_\sigma M \nonumber \\
    &= \beta \delta M_\text{Casimir},
\end{align}

which is expected, given the result in \eqref{quantum_statistical_relation}. The first law can then be written as\footnote{Recall that $M=M_0 + M_\text{BG} = M_0 + M_\text{Casimir}$.}

\begin{equation}
    dM = d M_0 + d M_\text{Casimir} = \delta M_\text{Casimir} +  T dS + \Omega_a dJ_a + \Omega_b dJ_b + \Phi dQ.
\end{equation}

We have already seen that the first term on the left-hand side equals the sum of all but the first term on the right-hand side. The Casimir energy is only a function of $a$ and $b$ and does not depend on $r_+$ or $q$. The term $d M_\text{Casimir}$ is hence just the variation of the Casimir energy with respect to $a$ and $b$. Using \eqref{delta_I=beta_delta_M} this term can be calculated from
\begin{equation} \label{dM_casimir}
    \delta M_\text{Casimir} = \beta^{-1} \int_{ [0, \,\beta] \times S^{3}} \text{d}^4x \sqrt{-h} \mathcal{A} \delta\sigma.
\end{equation}

\subsection{Extended Thermodynamics and Volume Calculation} \label{sec:Smarr_with_M}

Allowing the cosmological constant to vary leads to another treatment  called extended thermodynamics \cite{Kastor:2009wy, Kubiznak:2012wp}. In this treatment we are adding another pair of thermal quantities, namely the pressure, $P = {3g^2 / 4 \pi}$, and its conjugate thermodynamic volume
\begin{equation} \label{V}
    V = \left(\frac{\partial M}{\partial P}\right)_{S, Q, J_a, J_b}.
\end{equation}

This pair is supposed to obey Smarr's relation \eqref{Smarr} as in \cite{ Kubiznak:2012wp}, but it is straightforward to check that the volume in \eqref{V} does not satisfy this relation. In this section we analyze the reason behind this contradiction and show how it can be fixed. We are unaware of any similar attempts to address this issue for this or any other black hole solutions in the literature.

\par 

The relation in \eqref{V} can equivalently be written as

\begin{equation}
    V = \frac{\partial g}{\partial P} \left(\frac{\partial M}{\partial g}\right)_{S, Q, J_a, J_b}.
\end{equation}

It is easy to see that the last term on the RHS induces a variation in the boundary metric analogous to variations with respect to $a$ and $b$ in $\S$\ref{sec:verification_of_first_law}. We thus need to add a compensating term to the thermodynamic volume to account for this variation. Let us go back to the metric in \eqref{bg_metric_bar_coordinates}. We now switch to a new coordinate system $\left(t, \bar{r}, \bar{\theta}^{\prime\prime}, \phi^{\prime\prime}, \psi^{\prime\prime}\right)$, where

\begin{equation} \label{double_prime_coords}
    \tan^{2} \bar{\theta}=\left(1+\frac{\delta \Xi_{a}}{\Xi_{a}}-\frac{\delta \Xi_{b}}{\Xi_{b}}\right) \tan ^{2} \bar{\theta}^{\prime\prime}, \quad \phi=\phi^{\prime\prime}-a^2 t\delta g, \quad \psi=\psi^{\prime\prime}-b^2 t\delta g.
\end{equation}

We emphasize that the variations of $\Xi_a$ and $\Xi_b$ are now with respect to $g$, and the rotation parameters $a$ and $b$ are kept fixed throughout this subsection. With this in mind, it is easy to see that the variation of the metric follows the form

\begin{equation} \label{variation_of_bg_metric_wrt_g}
    d \bar{s}_\text{BG}^{2} \rightarrow\left(1-\frac{\delta \Xi_{a}}{\Xi_{a}} \sin ^{2} \bar{\theta}-\frac{\delta \Xi_{b}}{\Xi_{b}} \cos ^{2} \bar{\theta}\right) d \bar{s}_\text{BG}^{2}.
\end{equation}

Let us denote the Weyl factor in this section by $\delta \tilde{\sigma}$. We define an effective thermodynamic volume $\bar{V}$ that takes care of the extra terms in the mass that arise from the transformation \eqref{variation_of_bg_metric_wrt_g}  by
\begin{align} \label{V_bar}
    \bar{V} &=  \left(\frac{\partial M}{\partial P}\right)_{S, Q, J_a, J_b} + \frac{\partial (\delta_{\tilde{\sigma}} M)}{\partial P}.
\end{align}
This is a generalized definition of the volume that reduces to the usual definition in \eqref{V_0} and \eqref{V} in the absence of a conformal anomaly and still satisfies Smarr's formula in the presence of the latter. Furthermore, this definition satisfies Smarr's formula even if we use the ADM \cite{chen2005} or the Kounterterms method \cite{kounterterms_mass}. These two techniques give the mass expression in \eqref{M_0}, resulting in the last term on the RHS vanishing. Expression \eqref{V_bar} will hence give the same quantity in \eqref{V_0}, which we have already verified that it satisfies Smarr's formula if the mass is given by \eqref{M_0}. 

\par 

In AdS$_5$, it is straightforward to verify that transformation \eqref{variation_of_bg_metric_wrt_g} leads to a variation of the mass given by
\begin{align} \label{delta_I_sigma_tilde}
    \delta_{\tilde{\sigma}} M &= \beta^{-1} \delta_{\tilde{\sigma}} I_\text{ren} \nonumber \\
    &=  \beta^{-1} \int_{ [0, \,\beta] \times S^{3}} \text{d}^4x \sqrt{-h} \mathcal{A} \delta{\tilde{\sigma}} \nonumber \\
    &= \frac{\pi}{96 g^2 G} \left[ \frac{\delta \Xi _a (\Xi _a-\Xi _b) (\Xi _a+\Xi _b)}{\Xi _a{}^2 \Xi _b}-\frac{\delta \Xi _b (\Xi _a-\Xi _b) (\Xi _a+\Xi _b)}{\Xi _a \Xi _b{}^2}\right] \nonumber \\
    &= \frac{\pi}{96 g^2 G} \delta \left(\frac{\Xi_{a}}{\Xi_{b}}+\frac{\Xi_{b}}{\Xi_{a}}\right) .
\end{align}

The first line is just eq. \eqref{delta_I_in_A}, and we used relation \eqref{delta_I=beta_delta_M} to arrive at the second line. We now calculate the volume in \eqref{V_bar} for our solution:
\begin{align} \label{V_bar_calc}
    \bar{V} = V_0  -\frac{\pi ^2 \left[-9 g^2 \left(a^2+b^2\right)+g \left(a+7 a^2 b^2+b\right)+9\right]}{72 g \Xi _a \Xi _b}.
\end{align}

It is straightforward to show that the expression in \eqref{V_bar_calc} satisfies Smarr's formula with the mass calculated via the counterterms method,
\begin{align} \label{Smarr_for_with_V_bar}
    M = \frac{3}{2} \left(TS + \Omega_a J_a + \Omega_b J_b \right) + \Phi Q - P\bar{V}.
\end{align}

Alternatively, one can define the thermodynamic volume by \eqref{V} and add a compensating term to Smarr's formula, 
\begin{align} \label{Smarr_for_with_V}
    M = \frac{3}{2} \left(TS + \Omega_a J_a + \Omega_b J_b \right) + \Phi Q - \left(PV + P \frac{\partial (\delta_{\tilde{\sigma}} M)}{\partial P} \right).
\end{align}

\subsection{The First Law in Extended Thermodynamics}

We discuss here the first law in extended thermodynamics for the general black hole in five dimensions. In extended thermodynamics, the mass/energy $M$ is nothing but the enthalpy

\begin{equation}
H = U + P\bar{V}.
\end{equation}

The internal energy is thus found from

\begin{equation}
    U = M - P\bar{V}.
\end{equation}

With this, the first law in terms of $U$ can be written as

\begin{equation} \label{EPhS_first_law}
    dU = T dS + \sum_i \Omega_i dJ_i + \Phi dQ - Pd\bar{V}.
\end{equation}

It is easy to show that 
\begin{align} \label{EPhS_first_law_r,q}
    \frac{\partial U}{\partial r_+} &= T \frac{\partial S}{\partial r_+} + \sum_i \Omega_i \frac{\partial J_i}{\partial r_+} + \Phi \frac{\partial Q}{\partial r_+} - P \frac{\partial \bar{V}}{\partial r_+}, \nonumber \\
    \frac{\partial U}{\partial q} &= T \frac{\partial S}{\partial q} + \sum_i \Omega_i \frac{\partial J_i}{\partial q} + \Phi \frac{\partial Q}{\partial q} - P \frac{\partial \bar{V}}{\partial q},
    \nonumber \\
    \frac{\partial U}{\partial g} &= T \frac{\partial S}{\partial g} + \sum_i \Omega_i \frac{\partial J_i}{\partial g} + \Phi \frac{\partial Q}{\partial g} - P \frac{\partial \bar{V}}{\partial g}.
\end{align}

When we vary with respect to $c=a, b$ we find

\begin{equation} \label{EPhS_first_law_a,b}
    \frac{\partial U}{\partial c} = T \frac{\partial S}{\partial c} + \sum_i \Omega_i \frac{\partial J_i}{\partial c} + \Phi \frac{\partial Q}{\partial c} - P \frac{\partial \bar{V}}{\partial c} + \frac{\partial M_\text{BG}}{\partial c}.
\end{equation}

Eqs. \eqref{EPhS_first_law_r,q} and \eqref{EPhS_first_law_a,b} can be combined in the form

\begin{equation}
    dU = T dS + \sum_i \Omega_i dJ_i + \Phi dQ - Pd\bar{V} + \delta_\sigma M_\text{BG},
\end{equation}

where the variation $\delta_\sigma$ is understood to be w.r.t. to $a$ and $b$ but \textit{not} $g$.

\par 

This completes our investigation of the extended thermodynamics of
the general charged rotating black hole in five dimensions where we saw that Smarr's relation as well as the first law are satisfied using the counterterms method.

%% file: 7-Conclusion.tex
\section{Conclusion} \label{sec:conclusion}

We used the counterterms subtraction method to calculate various physical quantities of the charged rotating black holes in AdS$_5$ introduced in \cite{pope}. We showed that the resulting quantities satisfy the known thermodynamic relations in the cases of a varying and fixed cosmological constant, i.e., in extended and regular thermodynamics. All these quantities satisfy the Gibbs-Duhem relation, the first law, and Smarr's relation.

\par 

Quantum corrections break conformal symmetries in four-dimensional field theories, producing conformal anomalies. This classical symmetry on the boundary is realized as a subset of the bulk diffeomorphisms which is called the Penrose-Brown-Hennaux (PBH) transformation \cite{Penrose:1986ca,Brown:1986nw}. For boundary field theories with a non-vanishing conformal anomaly, a PBH transformation will not leave the gravitational on-shell action invariant. As was generally argued in \cite{skenderis}, this variation affects only the mass, but not the other quantities in the first law. This is because these other quantities can be written as integrals over the horizon. 

\par 

We showed that the mass calculated from the counterterms method can be written as a sum of two terms, one containing the mass parameter and electric charge, and another which is nothing but the background energy of the spacetime. The background energy is not restricted to the black hole region and its flux depends on the boundary. An important boundary condition for the AdS/CFT correspondence (or for the variation problem in \cite{skenderis}) is to keep the boundary metric fixed by utilizing a PBH or boundary Weyl transformation. As a result, one can use a Weyl transformation to cancel the variation in the mass term that results from varying the boundary metric. This leaves the first law satisfied.

\par

In this work we used the counterterms subtraction method to calculate the renormalized on-shell action for the general rotating charged AdS solution presented in \cite{pope}, as well as its mass and angular momenta. We showed that these quantities satisfy the first law and the Gibbs-Duhem relation. Going to extended thermodynamics \cite{Kastor:2009wy,Kubiznak:2012wp} where we allow the cosmological constant to vary and act as a pressure, a naive calculation of first law shows that it is not satisfied. We showed that a similar issue exists with the volume defined in the extended-thermodynamics treatment \cite{Kastor:2009wy,Kubiznak:2012wp}, meaning varying this quantity does not leave the boundary metric fixed and one needs to use a compensating PBH term to keep the volume fixed. This modification is important to satisfy the first law in this case as well as the generalized Smarr's formula when the counterterms subtraction method is used in the presence of a conformal anomaly. 

\par

We calculated the renormalized stress tensor and conformal anomaly of the CFT living on the boundary as predicted by the AdS/CFT duality, i.e., from the gravity side. We showed that these quantities coincide with the quantities calculated on the field theory side on a rotating Einstein Universe. Furthermore, we calculated the Casimir energy in the CFT and verified that it exactly matches the background energy of the bulk theory. 

\par

Finally, we showed that the calculation of the thermodynamic volume induces a Weyl transformation on the boundary metric, which adds extra terms to the calculated expression. We have generalized the definition of the thermodynamic volume in a way that accounts for these extra terms in the presence of a conformal anomaly. We have shown that our definition leads to a volume term that satisfies Smarr's relation as well as the first law in extended thermodynamics when the counterterms method is used. It would be interesting to use the quantities calculated here to study various phase transitions for the five-dimensional charged rotating solution \cite{pope}, which are expected to have a rich structure.

%% file: References.bib
@article{awad2007,
    title = {The First Law, Counterterms and Kerr-AdS$_5$ Black Holes},
   volume={18},
   number={03},
   journal={International Journal of Modern Physics D},
   publisher={World Scientific Pub Co Pte Lt},
   author={Awad, Adel M.},
   year={2009},
   month={3},
   pages={405–418}
}

@article{awad2016,
   title={Weyl Anomaly and Initial Singularity Crossing},
   volume={93},
   
   journal={Physical Review D},
   publisher={American Physical Society (APS)},
   author={Awad, Adel M.},
   year={2016},
   month={4}
}

@article{awad_johnsson1,
   title={Holographic Stress Tensors for Kerr-AdS Black Holes},
   volume={61},
  
   journal={Physical Review D},
   publisher={American Physical Society (APS)},
   author={Awad, Adel M. and Johnson, Clifford V.},
   year={2000},
   month={3}
}

@article{awad_johnsson3,
   title={Higher Dimensional Kerr-AdS Black Holes and the AdS/CFT Correspondence},
   volume={63},
   
   number={12},
   journal={Physical Review D},
   publisher={American Physical Society (APS)},
   author={Awad, Adel M. and Johnson, Clifford V.},
   year={2001},
   month={5}
}

@book{Penrose:1986ca,
    author = "Penrose, R. and Rindler, W.",
    title = "{Spinors and Space-time Vol. 2: Spinor and Twistor Methods in Space-time Geomtry}",
    publisher = "Cambridge University Press",
    series = "Cambridge Monographs on Mathematical Physics",
    month = "4",
    year = "1988"
    }

@article{Hawking:1982dh,
    author = "Hawking, S. W. and Page, Don N.",
    title = "{Thermodynamics of Black Holes in Anti-de Sitter Space}",
    reportNumber = "PRINT-83-0019 (CAMBRIDGE)",
    doi = "10.1007/BF01208266",
    journal = "Commun. Math. Phys.",
    volume = "87",
    pages = "577",
    year = "1983"
}

@article{Brown:1986nw,
    author = "Brown, J. David and Henneaux, M.",
    title = "{Central Charges in the Canonical Realization of Asymptotic Symmetries: An Example from Three-Dimensional Gravity}",
    journal = "Commun. Math. Phys.",
    volume = "104",
    pages = "207--226",
    year = "1986"
}

@book{birrel_davies,
    title = {Quantum Fields in Curved Space},
    author = {Birrell, N.D. and Davies, P.C.W.},
    isbn = {0 521 27858 9},
    year = {1984},
    publisher = {Cambridge University Press},
}

@article{boruch,
    author = "Boruch, Jan and Heydeman, Matthew T. and Iliesiu, Luca V. and Turiaci, Gustavo J.",
    title = "{BPS and Near-BPS Black Holes in AdS$_5$ and Their Spectrum in $\mathcal{N}=4$ SYM}",
    eprint = "2203.01331",
    archivePrefix = "arXiv",
    primaryClass = "hep-th",
    month = "3",
    year = "2022"
}

@article{brown_york,
   title={Quasilocal Energy and Conserved Charges Derived from the Gravitational Action},
   volume={47},
  
   number={4},
   journal={Physical Review D},
   publisher={American Physical Society (APS)},
   author={Brown, J. David and York, James W.},
   year={1993},
   month={2},
   pages={1407–1419}
}

@article{catastrophic_holography,
   title={Charged AdS Black Holes and Catastrophic Holography},
   volume={60},
 
   number={6},
   journal={Physical Review D},
   publisher={American Physical Society (APS)},
   author={Chamblin, Andrew and Emparan, Roberto and Johnson, Clifford V. and Myers, Robert C.},
   year={1999},
   month={8}
}

@article{chen2005,
    author = "Chen, W. and Lü, H. and Pope, C. N.",
    title = "{Mass of Rotating Black Holes in Gauged Supergravities}",
    eprint = "hep-th/0510081",
    archivePrefix = "arXiv",
    reportNumber = "MIFP-05-24",

    journal = "Phys. Rev. D",
    volume = "73",
    pages = "104036",
    year = "2006"
}

@article{gibbons2005,
   title={The First Law of Thermodynamics for Kerr–Anti-de Sitter Black Holes},
   volume={22},
  
   number={9},
   journal={Classical and Quantum Gravity},
   publisher={IOP Publishing},
   author={Gibbons, G. W. and Perry, M. J. and Pope, C. N.},
   year={2005},
   month={5},
   pages={1503–1526}
}

@article{grunau,
    author = "Grunau, Saskia and Neumann, Hendrik",
    title = "{Thermodynamics of a Rotating Black Hole in Minimal Five-Dimensional Gauged Supergravity}",
    eprint = "1502.06755",
    archivePrefix = "arXiv",
    primaryClass = "gr-qc",
   
    journal = "Class. Quant. Grav.",
    volume = "32",
    number = "17",
    pages = "175004",
    year = "2015"
}

@article{Kastor:2009wy,
    author = "Kastor, David and Ray, Sourya and Traschen, Jennie",
    title = "{Enthalpy and the Mechanics of AdS Black Holes}",
    eprint = "0904.2765",
    archivePrefix = "arXiv",
    primaryClass = "hep-th",
    doi = "10.1088/0264-9381/26/19/195011",
    journal = "Class. Quant. Grav.",
    volume = "26",
    pages = "195011",
    year = "2009"
}

@article{ Kubiznak:2012wp,
    author = "Kubiznak, David and Mann, Robert B.",
    title = "{P-V criticality of charged AdS black holes}",
    eprint = "1205.0559",
    archivePrefix = "arXiv",
    primaryClass = "hep-th",
    doi = "10.1007/JHEP07(2012)033",
    journal = "JHEP",
    volume = "07",
    pages = "033",
    year = "2012"
}

@article{holography_and_weyl_anomaly,
    author = "Henningson, Mans and Skenderis, Kostas",
    editor = "Lust, D. and Otto, H. J.",
    title = "{Holography and the Weyl Anomaly}",
    eprint = "hep-th/9812032",
    archivePrefix = "arXiv",
    reportNumber = "GOTEBORG-ITP-98-14, SPIN-1998-08, SPIN-1998-8",
   
    journal = "Fortsch. Phys.",
    volume = "48",
    pages = "125--128",
    year = "2000"
}

@article{kounterterms_mass,
	
	year = 2018,
	month = {6},
	publisher = {{IOP} Publishing},
	volume = {1043},
	pages = {012022},
	author = {Felipe Díaz-Martínez and Rodrigo Olea},
	title = {Conserved Quantities for a Charged Rotating Black Holes in 5D Einstein-Maxwell-Chern-Simons Theory},
	journal = {Journal of Physics: Conference Series},
}

@article{kraus,
  title={A Stress Tensor for Anti-de Sitter Gravity},
  author={Vijay Balasubramanian and Per Kraus},
  journal={Communications in Mathematical Physics},
  year={1999},
  volume={208},
  pages={413-428}
}

@article{large_n_lim, volume={38},
    title = {The Large N Limit of Superconformal Field Theories and Supergravity},
  
    number={4},
    journal={International Journal of Theoretical Physics},
    publisher={Springer Science and Business Media LLC},
    author={Maldacena, Juan},
    year={1999},
    pages={1113–1133}
}

@article{pope,
   title={General Nonextremal Rotating Black Holes in Minimal Five-Dimensional Gauged Supergravity},
   volume={95},
  
   number={16},
   journal={Physical Review Letters},
   publisher={American Physical Society (APS)},
   author={Chong, Z.-W. and Cvetič, M. and Lü, H. and Pope, C. N.},
   year={2005},
   month={10}
}

@article{pope2011,
  title={Black Hole Enthalpy and an Entropy Inequality for the Thermodynamic Volume},
  author={Cveti{\v{c}}, Mirjam and Gibbons, Gary W and Kubiz{\v{n}}{\'a}k, D and Pope, Christopher N},
  journal={Physical Review D},
  volume={84},
  number={2},
  pages={024037},
  year={2011},
  publisher={APS}
}

@article{skenderis,
   title={Thermodynamics of Asymptotically Locally AdS Spacetimes},
   volume={2005},
  
   number={08},
   journal={Journal of High Energy Physics},
   publisher={Springer Science and Business Media LLC},
   author={Papadimitriou, Ioannis and Skenderis, Kostas},
   year={2005},
   month={8},
   pages={004–004}
}

@article{silva2006,
    author = "Silva, Pedro J.",
    title = "{Phase Transitions and Statistical Mechanics for BPS Black Holes in AdS/CFT}",
    eprint = "hep-th/0610163",
    archivePrefix = "arXiv",
  
    journal = "JHEP",
    volume = "03",
    pages = "015",
    year = "2007"
}

@article{surface_terms,
   title={Surface Terms as Counterterms in the AdS/CFT Correspondence},
   volume={60},
  
   number={10},
   journal={Physical Review D},
   publisher={American Physical Society (APS)},
   author={Emparan, Roberto and Johnson, Clifford V. and Myers, Robert C.},
   year={1999},
   month={10}
}

@article{witten1998,
    title={Anti de Sitter Space and Holography},
    author={Edward Witten},
    year={1998},
    journal={Adv.Theor.Math.Phys.},
    eprint={hep-th/9802150},
    archivePrefix={arXiv},
    primaryClass={hep-th}
}
